\documentclass[preprint,3p,times,sort&compress]{elsarticle}

\usepackage{multirow,setspace,times,amssymb,amsmath,graphicx,color,rotating,subfigure,url}
\usepackage{lineno,color}
\usepackage{natbib}
\usepackage{booktabs}%

\usepackage[table]{xcolor}
\usepackage{tabularx}
\usepackage{graphicx} 
\usepackage{epstopdf}
\usepackage{mathrsfs}
\usepackage{makecell}
\usepackage[bookmarks=true,colorlinks,linkcolor=blue,anchorcolor=blue,citecolor=blue,unicode]{hyperref}
\usepackage{bookmark}

\hypersetup{CJKbookmarks=true}%
\journal{Elsevier}
\begin{document}

\begin{frontmatter}
\title{Robustness of the international oil trade network under targeted attacks to economies}

\author[SB]{Na Wei}
\author[SB,RCE]{Wen-Jie Xie\corref{coa}}
\ead{wjxie@ecust.edu.cn}
\author[SB,RCE,DM]{Wei-Xing Zhou}
\ead{wxzhou@ecust.edu.cn}
\cortext[coa]{Corresponding author. Corresponding to: 130 Meilong Road, P.O. Box 114, School of Business, East China University of Science and Technology, Shanghai 200237, China.}

\address[SB]{School of Business, East China University of Science and Technology, Shanghai 200237, China}
\address[RCE]{Research Center for Econophysics, East China University of Science and Technology, Shanghai 200237, China}
\address[DM]{Department of Mathematics, East China University of Science and Technology, Shanghai 200237, China}

\begin{abstract}
In the international oil trade network (iOTN), trade shocks triggered by extreme events may spread over the entire network along the trade links of the central economies and even lead to the collapse of the whole system. In this study, we focus on the concept of ``too central to fail'' and use traditional centrality indicators as strategic indicators for simulating attacks on economic nodes, and simulates various situations in which the structure and function of the global oil trade network are lost when the economies suffer extreme trade shocks. The simulation results show that the global oil trade system has become more vulnerable in recent years. The regional aggregation of oil trade is an essential source of iOTN's vulnerability. Maintaining global oil trade stability and security requires a focus on economies with greater influence within the network module of the iOTN. International organizations such as OPEC and OECD established more trade links around the world, but their influence on the iOTN is declining. We improve the framework of oil security and trade risk assessment based on the topological index of iOTN, and provide a reference for finding methods to maintain network robustness and trade stability.
\end{abstract}


\end{frontmatter}


\section{Introduction}

Oil as the primary fossil fuel is a vital strategic resource and widely used in the chemical industry, transportation, and other industries. The uneven distribution of oil on a global scale leads to the separation of supply and consumption markets. As a result, international oil trade becomes a critical complement to the balance of oil supply and demand \cite{Xi-Zhou-Gao-Liu-Zheng-Sun-2019-EnergyEcon,Caraiani-2019-EE, Sun-An-Gao-Guo-Wang-Liu-Wen-2019-Energy}.

Oil trade is not a simple point-to-point transaction, displaying the characteristics of a chain network structure across economies and regions \cite{Xie-Wei-Zhou-2020-EP}. Therefore, the oil trade system can be abstracted as an international oil trade network (iOTN),in which the economies are nodes and the trade relationships between them are edges \cite{Zhang-Lan-Xing-2018-IOP}. Fluctuations in oil consumption, supply, demand, storage, and price
can all cause changes in the global trade pattern and world economic situation \cite{Zhang-Ji-Fan-2015-EnergyEcon, Le-Chang-2013-EE, Rafiq-Sgro-Apergis-2016-EE}. It is of great significance for the government to know the development trend of oil trade, recognize the economy's trade status, and prevent oil supply risks \cite{An-Wang-Qu-Zhang-2018-Energy,Du-Wang-Dong-Tian-Liu-Wang-Fang-2017-AEn,Du-Dong-Wang-Zhao-Zhang-Vilela-Stanley-2019-Energy}.

Oil trade promotes energy cooperation and economic development. However, the iOTN as a complex system also faces shocks and challenges. In the closely connected iOTN, not only oil resources but the trade shocks and supply risks can be transmitted through trade relationships between economies. In 2014, the sharp increase in shale oil production in the United States caused the U.S.-Africa oil trade volume to plummet to its lowest point in 40 years, and Africa became the worst hit economy. The plunge in the U.S.-Africa oil trade has not only disrupted the most important economic ties between the U.S. and Africa, but has also had a ripple effect globally. Many economies suffered from potential trade shocks brought about by changes in oil imports of the U.S. This phenomenon of risk transmission and diffusion can be attributed to the cascade effect of network \cite{Kim-Eisenberg-Chun-Park-2017-PhysicaA,Wang-Chen-2008-PhysRevE}. The above facts show that when unexpected events  impact trade activities, the structure and functions of the iOTN will be affected and the global oil security may be endangered. Therefore, it is of great practical significance to integrate the study of trade risk transmission mechanism based on complex network theory into the framework of energy security assessment.

In recent years, focusing on economic fragility caused by factors such as oil supply interruption and political instability, there has been a growing literature on global oil security \cite{Brown-Huntington-2013-EnergyEcon,Cherp-Jewell-2014-EnergyPolicy,Ang-Choong-Ng-2015-RenewSustEnergRev}. A large body of literature provides multi-angle, multi-dimensional refinements for the energy security assessment framework \cite{Sovacool-2011-EnergyPolicy,Sovacool-Mukherjee-2011-Energy}. However, there is little literature on integrating the topological indicators of the iOTN into the traditional oil security assessment framework \cite{Liu-Cao-Liu-Shi-Cheng-Liu-2020-Energy}. How the roles and status of economies in the iOTN affect oil security and trade stability is still an open problem worth exploring.

Considering the oil security and trade structural stability implies the need to explore the stability of the iOTN against extreme events or the robustness when encountering cascade failures. In other words, the iOTN needs to have the ability to maintain certain structural integrity and functions when the economies suffer from terrorist attacks, foreign policy changes, and the energy trade channels between economies are blocked or interrupted. Therefore, robustness is an essential dynamic characteristic of the network system. The higher the robustness of a network, the more stable it is \cite{Zhou-Wang-Hang-2019-TRPE}.

The international oil market is turbulent, and the political stance of the economies and international relations are complicated. It is difficult to capture and detect the real changes of the oil trading system and the cascade effect after the impact, and it is even more difficult to conduct quantitative analysis and research. Firstly, we construct iOTNs based on international oil trade data sets. Secondly, we use the numerical simulation method to simulate attacks on the iOTNs and analyze the network's robustness changes under various targeted attacks. Finally, through the comparative analysis of different attacks, we find the effective way to maintain network robustness and trade stability, and provide suggestions for oil trade policy formulation and market supervision.

This article is organized as follows: Section \ref{S1:LitRev} reviews the literature; Section \ref{S1:Data:Methodology} introduces research data and methods; In section \ref{S1:EmpAnal}, we construct the international oil trade networks and uses the related method to conduct the empirical analysis; Section \ref{S1:Conclude} is the discussion and application.

\section{Literature review}
\label{S1:LitRev}

According to the resource dependence theory, close trade relations are needed between economies to exchange resources to meet their needs. Those complex relations often form a trade network pattern and promote the process of trade globalization. Network science become an essential method for analyzing the pattern of trade network \cite{Serrano-Boguna-2003-PRE,Fagiolo-Reyes-Schiavo-2009-PRE,An-Zhong-Chen-Li-Gao-2014-Energy}.

As the most crucial energy resource, oil's global trade always attracts the attention of researchers. The international oil trade system is evolving into a stable, integrated, and more efficient system \cite{An-Zhong-Chen-Li-Gao-2014-Energy,Xie-Wei-Zhou-2020-EP}.Economies are more interconnected, but global competition for oil resources has intensified. The addition of industrialized economies such as China and India increase the uncertainty of the development of oil trade \cite{Yu-Jessie-Sharmistha-2015-Energy}. In addition to factors such as supply and demand, technological advances and energy efficiency, geographic location, and competition and dependence between economies have increasingly prominent influence on oil trade \cite{Zhang-Ji-Fan-2015-EnergyEcon,Kharrazi-Fath-2016-EP,Kitamura-Managi-2017-AEn}. The researches about the iOTN at the system and individual levels reveal the general characteristics of oil trade and deepen the understanding of the oil trade system \cite{Le-Chang-2013-EE,Rafiq-Sgro-Apergis-2016-EE}. The researches also have important practical guiding significance for the position recognition and trade strategy formulation of the economies in the iOTN.

The complexity and vulnerability of the oil trade system still leaves many issues to be explored and resolved \cite{Ji-Zhang-Fan-2014-ECM}. Therefore, the stability of the iOTN and trade risks are closely related to energy security and the stable development of global economy, and are hot research issues. The risk and instability arise from the chain impact of the failure of economies through the trade relationships in the iOTN, which can also be called cascade failure. When the cascade shock reaches a certain level and scope, it will cause the outbreak of systemic risks in the iOTN. It may eventually cause the collapse of the entire system and have a continuous impact on the global economy \cite{Battiston-Farmer-Flache-Garlaschelli-Haldane-Heesterbeek-Hommes-Jaeger-May-Scheffer-2016-Science}. Then, the stability of the iOTN is crucial to the development of economies and the global oil trade market \cite{Zhong-An-Gao-Sun-2014-PA,Sun-Gao-Zhong-Liu-2017-PA}.

From the concepts of ``too big to fail" and ``too connected to fail" to the concept of ``too central to fail", numerous researchers have paid attention to these concepts and have discussed them extensively. In this article, we focus on ``too central to fail" \cite{Yun-Jeong-Park-2019-JEconBehavOrgan,Zhong-An-Shen-Fang-Gao-Dong-2017-EP}. Central economies are very important for maintaining the structural stability and play a bridge role in the formation of trade networks \cite{Ji-Zhang-Fan-2014-ECM}. Traditional measures of network node importance or influence, such as degree, betweenness, and closeness, can be used to determine the economies with dominant positions \cite{Dablander-Hinne-2019-SR,Richmond-2019-JF}. Based on these indicators, many researchers propose new methods to measure the influence of network nodes \cite{Battiston-Puliga-Kaushik-Tasca-Caldarelli-2012-SR,Du-Wang-Dong-Tian-Liu-Wang-Fang-2017-AEn,Zhong-An-Shen-Fang-Gao-Dong-2017-EP}.

Different from the methods of directly selecting or designing node influence measures to identify the most influential nodes in previous studies \cite{Battiston-Puliga-Kaushik-Tasca-Caldarelli-2012-SR,Du-Wang-Dong-Tian-Liu-Wang-Fang-2017-AEn},
this article selects a series of classic node influence measures, and uses them as strategic indicators for simulated attacks on the economic nodes in the iOTN. This article  explores the changes in the stability of the iOTN structure when important economies are impacted, and compare those indicators. The findings can better the understanding of the role of economies with different critical positions in the transmission of oil trade risks.

\section{Data and Methodology}
\label{S1:Data:Methodology}

\subsection{International oil trade data}

The oil trade data is downloaded from the UN Comtrade database, and the HS code of data is 270900. The data comes from official data reported by both trade parties. We chose the import data with broader coverage and better integrity \cite{Fan-Ren-Cai-Cui-2014-EM}, which contains the oil import and export trade values of 256 economies from 1988 to 2017. We removed a small amount of economic trade data whose reporting source cannot be identified, such as ``Other Asia, nes". The deleted data is small in amount and will not affect the research results.

\subsection{Construction of iOTNs}

The global oil trade system can be abstracted as an international oil trade network (iOTN). We define the unweighted directed iOTN that can be represented by the $N\times N$ adjacency matrix $A=(a_{ij})$, where $N$ is the number of economies in the iOTN. The matrix element $a_{ij}=1$ indicates that there is an export trade from $i$ to $j$. We construct the yearly iOTNs from 1988 to 2017.

\begin{figure}[!t]
\centering
\includegraphics[width=0.95\linewidth]{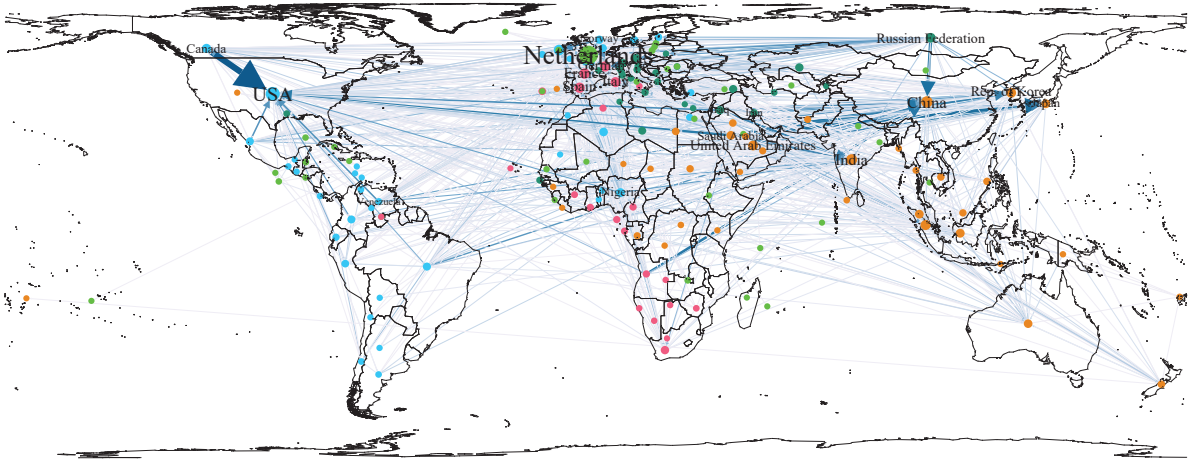}
\caption{The international oil trade network in 2017. Each node represents an economy, and a directed edge represents the oil trade relationship from one economy to another. The size of a node implies the number of the economy's partners. Different colors of the nodes indicate that the economies belong to different modules.}
\label{Fig:iOTN:Role:2017}
\end{figure}

Figure~\ref{Fig:iOTN:Role:2017} shows the iOTN in 2017. The nodes are economies, and the edges with direction indicate the oil trade relationships. The sizes of nodes are set according to the number of the economies' partners. The more the trade partners of the economy, the larger the node size. The different colors of the nodes indicate that the economies belong to different modules. The method of module division will be explained in the method section.

\subsection{Models of attack to economies in iOTN}

The scale-free characteristic of the iOTN enables the trade system to resist random attacks but is highly vulnerable to targeted attacks \cite{Liu-Cao-Liu-Shi-Cheng-Liu-2020-Energy}. Previous studies prove that attacks on a single important node can trigger large-scale cascade failures \cite{Crucitti-Latora-Marchiori-2004-PRE}.  A targeted attack model of the economies needs to be constructed to explore the structural stability of the iOTN in different attack situations.

A targeted attack involves a specific rule or indicator. We take the influence measures as the strategic indicators for the attacks. Simulating the destruction of the network under targeted attacks can serve as a prediction for the crisis of system collapse. Knowing under which specific circumstances the system is more prone to collapse can be a preparation for the reinforcement and protection of the subsequent system.

Based on these strategies and indicators, we select the economies with the higher importance rankings, accounting for $q$ of the total number of economies in the iOTN, to attack. Then, the selected economic nodes can be removed from the network. By changing the selection basis and value of $q$, we can simulate network structure damage under different attack strategies and degrees.

\subsection{Attack strategies based on the influence of economies}

The importance of economies usually depends on the analysis of node centrality indicators \cite{Dablander-Hinne-2019-SR}. Different centrality indicators reveal the influence of nodes in the network from different perspectives. Considering the types and scopes of the network topological connection, we choose three types of influence measures that were commonly used in previous studies \cite{Liu-Cao-Liu-Shi-Cheng-Liu-2020-Energy,Xie-Wei-Zhou-2020-EP}. Then, we use them as strategic indicators for subsequent economic attacks. The impacts of critical economies identified by traditional methods on the stability of the iOTN can be measured, and different measurement methods can be horizontally compared.

\subsubsection{Economic influence measures based on the local structure of the iOTN}

The influence of economies based on the local structure is mainly measured according to the local topological characteristics of iOTN such as degree centrality, local clustering coefficient. Some other methods are also derived according to the different local network structure and scope \cite{Chen-Lu-Shang-Zhang-Zhou-2012-PhysicaA}. We select the most used degree and local clustering coefficient as strategic indicators.

We define \textbf{indegree} $k_{\rm{in}}$ and \textbf{outdegree} $k_{\rm{in}}$ of economy $i$, which represent the number of import and export relationships.
\begin{equation}\label{Eq:Centrality:Degree}
K_{i}^{\rm{in}}=\sum_{j=1}^{N}a_{ji} ~~~ and ~~~ K_{i}^{\rm{out}}=\sum_{j=1}^{N}a_{ij},
\end{equation}
where $N$ is the total number of economies in the iOTN.

\textbf{Local clustering coefficient} $C_{\rm{c}}$ of the economy $i$ is ratio of the number of actual trade relationships between its neighboring economies to all possible trade relationships \cite{Watts-Strogatz-1998-Nature}. It is defined as:
\begin{equation}\label{Eq:Centrality:Clustering}
C_{\rm{c}}\left(i\right)=\frac{|\{a_{jk}:i,j \in N_{i},a_{ij} \in A\}|}{K_i(K_i-1)},
\end{equation}
where $N_i$ is the set of neighboring economies of economy $i$, $K_i=K_{i}^{\rm{in}}+K_{i}^{\rm{out}}$.

\subsubsection{Economic influence measures based on the global structure of the iOTN}

The node's influence is measured according to the network's global information. The most used methods include the shortest path (betweenness and closeness) and random walk (such as PageRank).

\textbf{Betweenness} $B_t(i)$ refers to the ratio of the number of shortest paths passing through a specific economy $i$ to the total number of shortest paths between any two economies \cite{Freeman-1977-Sociometry}:
\begin{equation}\label{Eq:Centrality:Betweenness}
B_t(i)=\sum_{st}\frac{n_{st}^i}{g_{st}},
\end{equation}
where $g_{st}$ is the total number of shortest paths from node $s$ to node $t$ and $n_{st}^i$ is the number of those paths pass through.

\textbf{Closeness} $C_{\rm{out}}$ considers the average length of the shortest path from each economy to other economies \cite{Hage-Harary-1995-SocNetworks}.
\begin{equation}\label{Eq:Centrality:Closeness}
C_{\rm{out}}\left(i\right)=\left(\frac{D_i}{N-1}\right)^2\frac1{C_i},
\end{equation}
where $D_{i}$ is the number of economies that can be reached from economy $i$ (excluding $i$), $N$ is the number of economies in the iOTN. $C_i$ is the sum of the distances from economy $i$ to all reachable economies, and for isolated nodes, $C_{out}(i)=0$. Considering that the iOTN is a directed network, Eq.~(\ref{Eq:Centrality:Closeness}) calculates outcloseness. Taking the trade directions into consideration, incloseness can also be calculated based on the Eq.~(\ref{Eq:Centrality:Closeness}). Incloseness is the reciprocal of the sum of the distances from all economies to the economy $i$.

\textbf{PageRank} is an indicator for measuring the influence of economies in the directed network \cite{Page-1999-StanfordInfoLab}. When calculating PageRank, the number of iterations is set to 100.

\textbf{Authorities and Hubs} are two indicators involved in the HITS algorithm, which were first proposed by Kleinberg,  and are inseparable in measuring the centrality \cite{Kleinberg-1999-JACM}. An economy with a high authority score is defined as the economy pointed by many economies with high hub centrality. Correspondingly, an economy with a high hubs score is defined as it points to many economies with high authority scores. The HITS algorithm is exquisite and can provide more information for node centrality in theory, so the index is relatively complicated. In this article, the hub or authority scores are converted according to the proportion of the economy in all economies, and the sum of the scores (hub or authority) is equal to 1.

\subsubsection{Economic influence measures based on modular structure of the iOTN}

Economies in the iOTN with similar functions have similar topological properties \cite{Guimera-Amaral-2005-JSM}. Therefore, there is often a modular structure. The connections between members within each module are relatively close, while the connections between the modules are relatively sparse \cite{Zhang-Lan-Xing-2018-IOP,Giudici-Huang-Spelta-2019-ESs}. The economic influence index based on the module structure considers the individual characteristics of the economy and mines the group information based on the network module.

We apply the module division method based on network structure to identify the critical position of each economy in the iOTN \cite{Guimera-Amaral-2005-Nature}. Firstly, the network needs to be divided into modules. Two indicators of within-module degree and participation coefficient can be proposed to measure the influence of economies on the internal and external modules where they are located. We use the classic modular algorithm to divide the iOTN into modules \cite{Newman-Girvan-2004-PRE,Newman-2006-PNAS,Blondel-Guillaume-Lambiotte-Lefebvre-2008-JSM}.

\textbf{Within-module degree} $Z_i$ measures how well-connected economy $i$ is to other nodes in its module \cite{Guimera-Amaral-2005-Nature}.
\begin{equation}\label{Eq:Modular:Economies:in}
Z_i=\frac{k_{i,s}-{\overline k}_{i,s}}{\sigma_{s}},
\end{equation}
where $k_{i,s}$ is the number of trade relationships between economy $i$ and other economies in module $s$. ${\overline k}_{i,s}$ is the average value of $k_{i,s}$ for economies in the module $s$. $\sigma_{s}$ is the standard deviation of $k_{i,s}$ in $s$. If the economy $i$ has trade relations with many economies in its module $s$, the value of $Z_i$ will be larger, implying that the trade changes of the economy $i$ can affect other economies in the module to a greater extent.

\textbf{Participation coefficient} $P$ measures the uniformity of connections between the economy $i$ and the economies in different modules \cite{Guimera-Amaral-2005-Nature}.
\begin{equation}\label{Eq:Modular:Economies:global}
P_i=1-\sum_{s=1}^{N_M}\left(\frac{k_{i,s}}{K_i}\right)^2,
\end{equation}
where $k_{i,s}$ is the number of relationships of economy $i$ to economies in module $s$, and $K_i$ is the total degree of economy $i$. $N_M$ is the total number of modules. The $P$ of an economy is close to 1 if its trade partners are uniformly distributed among all the modules. Furthermore, if its partners are within its module, $P=0$.

$P$ cannot measure the strength of the economy's influence in oil trade outside its module. Therefore, we introduce the outside-module degree \cite{Xu-Pan-Muscoloni-Xia-Cannistraci-2020-NC}.

\textbf{Outside-module degree} of economy $i$ measures how well-connected $i$ is with ports outside its own module \cite{Xu-Pan-Muscoloni-Xia-Cannistraci-2020-NC}. It is defined as:
\begin{equation}\label{Eq:Modular:Economies:out}
B_i=\frac{m_{i,s}-{\overline m}_{i,s}}{\sigma_{s}},
\end{equation}
where $m_{i,s}$ is the number of connections of economy $i$ to other economies out of its own
module $s$. ${\overline m}_{i,s}$ and $\sigma_{s}$
are respectively the average and standard deviation of $m_{i,s}$.

\subsection{Robustness of the iOTN}

To measure the function and structural integrity of the iOTN after an attack, we use the fraction of economies in the giant connected component (GCC) after removing a ratio of $q$ economies as the measure indicator \cite{Schneider-Moreira-Andrade-Havlin-Herrmann-2011-PNAS}, which can be defined as $S(q)$. Considering the various situations where the network suffers enormous damage but does not completely collapse, we introduce robustness measure $R$ to measure the function and structural integrity of the iOTN under different attack strategies \cite{Schneider-Moreira-Andrade-Havlin-Herrmann-2011-PNAS}:
\begin{equation}\label{Eq:Robustness:Measure}
R=\frac{1}{N}\sum_{n=1}^{N}S\left(\frac{n}{N}\right),
\end{equation}
where $N$ is the total number of economies in the iOTN, and $n=qN$. The normalization factor $\frac{1}{N}$ ensures that the robustness of networks with different sizes is comparable.

\section{Empirical analysis}
\label{S1:EmpAnal}

\subsection{Ranking of economic influence in the iOTN}

\begin{table*}[!t]
  \centering
  \caption{Top 10 economies identified based on different influence indicators in 2017}
    \scalebox{0.7}{
    \begin{tabular}{ccccccc}
    \toprule
    Rank & Pagerank & Outegree & Indegree & Outcloseness & Incloseness & {Betweenness} \\
    \midrule
    1   & Netherlands & USA & Netherlands & USA & Netherlands & Netherlands \\
    2   & India & Russian Federation & USA & Russian Federation & USA & USA \\
    3   & Spain & United Kingdom & India & United Kingdom & India & United Kingdom \\
    4   & USA & Nigeria & China & Nigeria & Spain & {United Arab Emirates} \\
    5   & France & Saudi Arabia & Spain & Saudi Arabia & France & India \\
    6   & Italy & Kazakhstan & Rep. of Korea & Algeria & Italy & China \\
    7   & Singapore & Iraq & Singapore & United Arab Emirates & Canada & Germany \\
    8   & Rep. of Korea & United Arab Emirates & France & Iraq & Germany & France \\
    9   & Canada & Algeria & Italy & Norway & Sweden & South Africa \\
    10  & Germany & Germany & Germany & Azerbaijan & United Arab Emirates & Azerbaijan \\
    \midrule
    Rank & Authorities & Hubs & Clustering & Within-module   & Outside-module   & Participation \\
    \midrule
    1   & Netherlands & USA & Saudi Arabia & Netherlands & Netherlands & Netherlands \\
    2   & China & Russian Federation & Norway & USA & USA & Israel \\
    3   & USA & Saudi Arabia & Venezuela & China & China & USA \\
    4   & Spain & Nigeria & Angola & Russian Federation & Russian Federation & Finland \\
    5   & India & United Kingdom & Papua New Guinea & South Africa & India & El Salvador \\
    6   & France & Iraq & Kuwait & India & South Africa & Mozambique \\
    7   & Italy & Algeria & Iraq & United Arab Emirates & Italy & Zambia \\
    8   & Rep. of Korea & Kazakhstan & Brazil & France & United Kingdom & Italy \\
    9   & Singapore & Libya & Chad & Spain & Rep. of Korea & Sweden \\
    10  & Germany & Norway & Canada & Rep. of Korea & Singapore & Libya \\
    \bottomrule
    \end{tabular}}%
  \label{tab:2017:node:criticalityRank10}%
\end{table*}%

The indicators to measure the influence of nodes in the network are applied to the real trade data set \cite{Dablander-Hinne-2019-SR,Richmond-2019-JF}. The critical nodes in the network may vary depending on the indicators and network types. However, there is relatively little literature on comparison and analysis of these indicators and methods themselves in the  study of the trade risk transmission mechanism under the framework of energy security,

We calculate the oil trade influence of economies in different years based on the economic influence measures proposed in Section~\ref{S1:Data:Methodology}. The trade influence of economies is continually changing over time. Taking the iOTN in 2017 as an example, we select the top 10 economies with the most considerable trade influence under different measures as shown in Table~\ref{tab:2017:node:criticalityRank10}.

In Table~\ref{tab:2017:node:criticalityRank10}, the ranking results under different measures are indeed different. The USA and the Netherlands both have a strong influence on the network's local, global, and community structure. In 2017, the Netherlands surpassed the USA in most centrality measures and was considered as the most critical economy. The surpass may be related to the expansion of both port oil trade in the Netherlands and the American energy transformation in recent years. Due to complex geopolitical relations and continuous regional conflicts, the dependence of the USA on traditional oil-producing regions in the Middle East and Africa has been significantly reduced. The USA has gradually moved from the largest oil importer to energy self-sufficiency \cite{An-Wang-Qu-Zhang-2018-Energy}. Emerging economies such as China and India have also demonstrated their importance in the iOTN.

\subsection{Correlation analysis between indicators of the economic influence}

Although different influence indicators measure the influence of economies in the iOTN from different aspects, the settings of the centrality indicators are all based on the topological structure of the iOTN, and there may be correlation between those indicators. We calculate the correlation between the 12 indicators and the results are shown in Table~\ref{tab:node:centrality:corr}.

\begin{table*}[!t]
  \centering
  \small
  \caption{Correlation of the influence indicators of oil trade economies in 2017.}
      \scalebox{0.9}{
    \begin{tabular}{l@{}llllll}
    \toprule
        & Pagerank & Outdegree & Indegree & Outcloseness & Incloseness & Authorities \\
    \midrule
    Pagerank & 1.000  &     &     &     &     &    \\
    Outdegree & 0.274$^{***}$  & 1.000  &     &     &     &    \\
    Indegree & 0.930$^{***}$  & 0.378$^{***}$  & 1.000  &     &     &  \\
    Outcloseness & 0.198$^{**}$  & 0.714$^{***}$  & 0.273$^{***}$  & 1.000  &     &  \\
    Incloseness & 0.521$^{***}$  & 0.211$^{**}$  & 0.607$^{***}$  & 0.065  & 1.000  &   \\
    Authorities & 0.748$^{***}$  & 0.434$^{***}$  & 0.929$^{***}$  & 0.315$^{***}$  & 0.677$^{***}$  & 1.000  \\
    Hubs & 0.214$^{**}$  & 0.959$^{***}$  & 0.293$^{***}$  & 0.771$^{***}$  & 0.146  & 0.343$^{***}$ \\
    Betweenness & 0.919$^{***}$  & 0.417$^{***}$  & 0.856$^{***}$  & 0.262$^{***}$  & 0.396$^{***}$  & 0.658$^{***}$ \\
    Clustering & 0.006  & 0.617$^{***}$  & 0.082  & 0.549$^{***}$  & 0.003  & 0.152$^{*}$  \\
    Within-module   & 0.671$^{***}$  & 0.659$^{***}$  & 0.792$^{***}$  & 0.485$^{***}$  & 0.483$^{***}$  & 0.791$^{***}$ \\
    Outside-module   & 0.650$^{***}$  & 0.723$^{***}$  & 0.767$^{***}$  & 0.520$^{***}$  & 0.393$^{***}$  & 0.758$^{***}$ \\
    Participation   & 0.252$^{***}$  & 0.489$^{***}$  & 0.335$^{***}$  & 0.501$^{***}$  & 0.494$^{***}$  & 0.408$^{***}$ \\
    \toprule
        & Hubs & Betweenness & Clustering & Within-module & Outside-module & Participation \\
    \midrule
    Hubs & 1.000  &     &     &     &     &  \\
    Betweenness  & 0.315$^{***}$  & 1.000  &     &     &     &  \\
    Clustering & 0.704$^{***}$  & 0.039  & 1.000  &     &     &  \\
    Within-module & 0.550$^{***}$  & 0.686$^{***}$  & 0.247$^{**}$  & 1.000  &     &  \\
    Outside-module & 0.629$^{***}$  & 0.689$^{***}$  & 0.299$^{***}$  & 0.884$^{***}$  & 1.000  &  \\
    Participation   & 0.518$^{***}$  & 0.209$^{**}$  & 0.391$^{***}$  & 0.301$^{***}$  & 0.409$^{***}$  & 1.000  \\
    \bottomrule
    $^{*}p<0.05,^{**}p<0.01,^{***}p<0.001.$\\
    \end{tabular}}%
  \label{tab:node:centrality:corr}%
\end{table*}%

Table~\ref{tab:node:centrality:corr} shows the Spearman correlation coefficients between the influence measures. When measuring the trade influence of economies, most of the indicators have a significant correlation. The outdegree and hubs have a significant and highly positive correlation, and the correlation coefficient is 0.959. The correlation coefficients of indegree and PageRank, and of indegree and authority are 0.930 and 0.929 respectively.

Although there are close relationships between the economic influence measures, their calculation basis is different and cannot replace each other. The optimal influence measure needs to be selected according to specific issues. This study aims to find the economies that have a more significant impact on the stability of the oil trade network structure. Therefore, the horizontal comparison of the influence measures is of great significance for selecting the key economies that deserves more attention and maintaining the structural stability of the iOTN.

\subsection{Analysis on the oil trade influence of international organizations}

\begin{figure}[!t]
\centering
\includegraphics[width=0.95\linewidth]{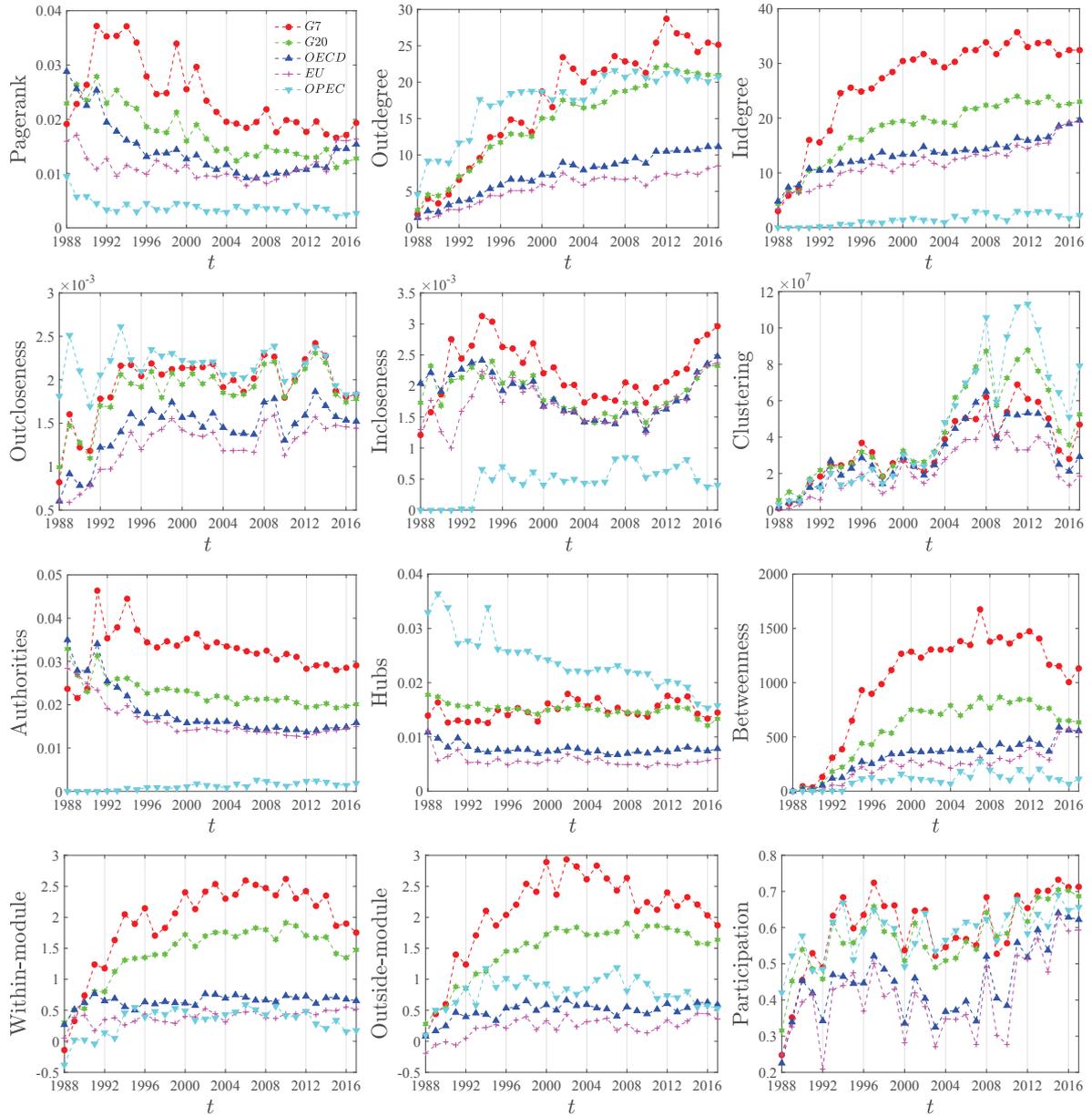}
\caption{The evolution of the oil trade influence of international organizations from 1988 to 2017.} \label{Fig:iOTN:Organization}
\end{figure}

In the global energy governance system, organizations such as the Organization of Petroleum Exporting Countries (OPEC) and the Organization for Economic Cooperation and Development (OECD) are important international economic institutions, performing an important role in global economic governance and global economic order coordination. In the 21st century, emerging economies have grown rapidly, and the global economic landscape has continuously changed. Exploring the changes of energy trade influence of different international economic organizations can deepen the understanding of the changes in the energy market patterns  and the international situation.

We take several typical international organizations as the research objects, and take the average influence value of the economies included in these organizations as their oil trade influence. Fig.~\ref{Fig:iOTN:Organization} shows the trade influence evolution of the OPEC, OECD, European Union (EU), G7, and G20 based on different influence indicators.

It can be seen from Fig.~\ref{Fig:iOTN:Organization} that the oil trade influence of international organizations has a relatively apparent trend under most measures. The influence measured by the network's global structure, such as PageRank, authority, and hubs, has a clear downward trend. The influence based on the local structure, such as degree and local clustering coefficient, shows an upward trend. The influence based on the modular structure was an upward trend in the early stage and then gradually stabilized and fluctuated slightly. The influence with the participation coefficient as an indicator fluctuates significantly with the year.

It can be concluded that as more and more economies join the process of trade globalization, international organizations have developed closer trade ties with neighboring economies, and their influence on a local scale is increasing. However, as far as the global situation is concerned, the international status, trade policies and oil trade patterns of economies are constantly changing.  The impact of a single international organization on global oil trade is declining.

\begin{figure}[!t]
\centering
\includegraphics[width=0.95\linewidth]{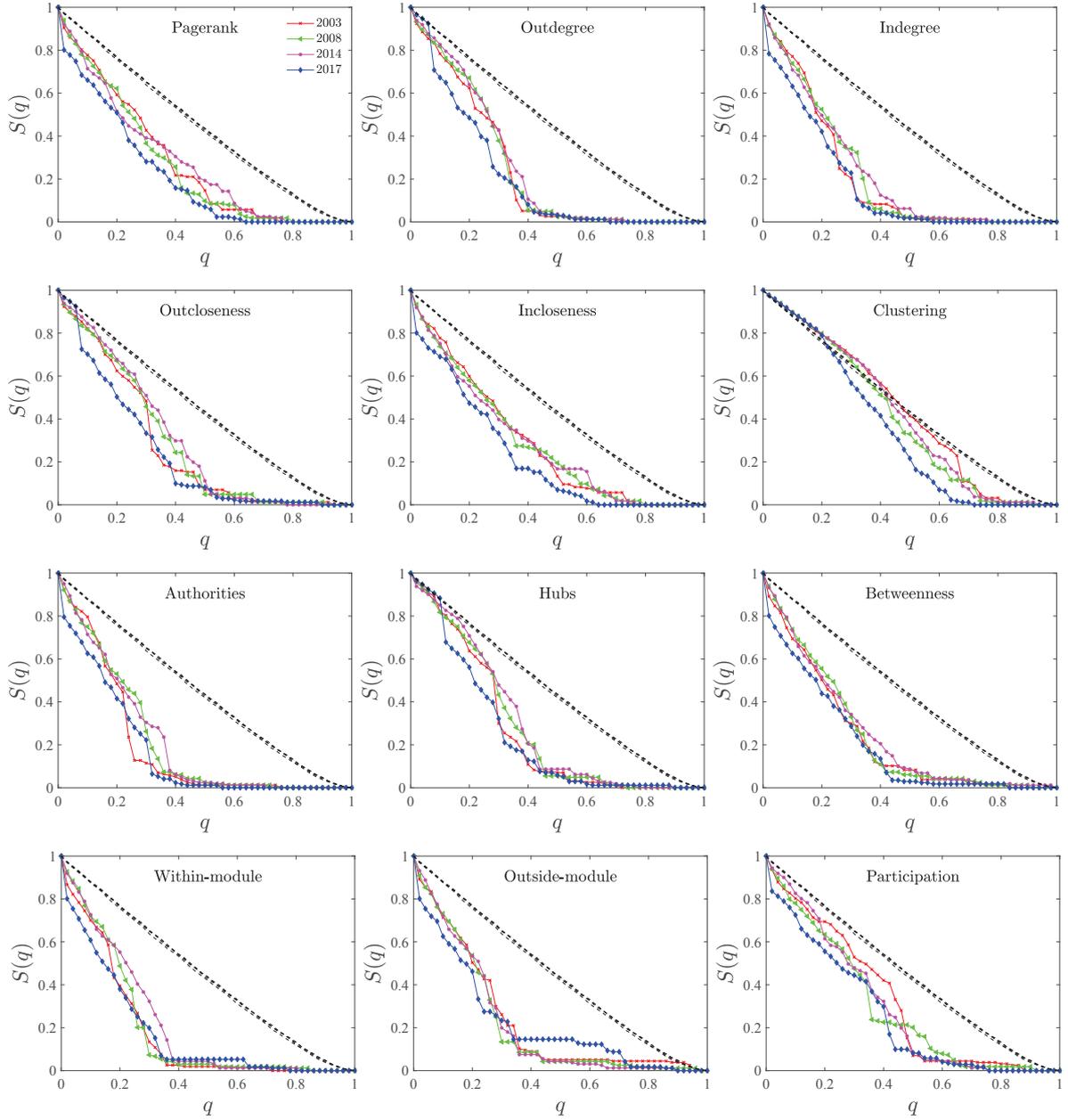}
\caption{The structural integrity of the iOTN under attacks of economic nodes.}
\label{Fig:iOTN:CentrRoleAttack}
\end{figure}

The nature of international organizations and the characteristics of their membership determine the trend of oil trade influence. After the first oil crisis hit Western economies, the G7 was born to respond to the global economic crisis and to stabilize the international economic system. The outbreak of the Asian financial crisis in 1997 highlighted the importance of developing countries. With the participation of major developing countries, the G20 came into being. Therefore, the emergence of the G7 and the G20 are closely related to the economic crisis. Similarly, the EU and the
OECD aim to promote international cooperation and joint economic development, and they have many common members. The trade influence measured by the more relevant centrality indicators has the same trend. These facts explain the similar trends in the oil trade influence of different organizations and indicators.

OPEC has certain peculiarities among several organizations, being the earliest established and the most influential producer and exporter of raw materials. Its purpose is to coordinate and unify oil policies, maintain the stability of international oil market prices, and safeguard oil-producing countries' interests. Therefore, its impact on oil trade is not limited to its trade, but it also balances world power by affecting oil prices. With the strengthening of international cooperation, the trade cooperation of various economies is more extensive.However,
based on the oil trade network's topological structure, OPEC's trade influence has not been significantly improved from a multidimensional perspective. Its global trade influence based on trade export capacity as measured by the hubs indicator is declining. This result is also consistent with OPEC's declining impact on global oil production mentioned in previous studies \cite{AlRousan-Sbia-Tas-2018-EnergyEcon}.

\subsection{Analysis on the ability of the iOTN to resist risks under attacks}

The oil trade of the economies is continually diversifying and the structure of the iOTN is becoming increasingly complex. Trade shocks and risks caused by unexpected events such as oil supply interruptions, wars, economic sanctions may pass along intricate trade links. From the perspective of network structure, when the iOTN suffers from trade shocks (random attacks or targeted attacks), it needs to be robust enough to maintain trade stability and sustainable global economic development. To measure the robustness of the iOTN under economic attacks, which can also be called the anti-risk ability of the network structure, we use the influence measures of economies as strategic indicators of node attacks. The changes in the integrity of the network structure after the network is attacked are shown in Fig.~\ref{Fig:iOTN:CentrRoleAttack}.

In Fig.~\ref{Fig:iOTN:CentrRoleAttack}, we show the results of simulated attacks in 2003, 2008, 2014 and 2017. For comparison, the black dashed lines represent the results of randomly removing economies with a ratio of $q$. Under the random attack, the GCC size decreases linearly as the attack intensity increases. Under the targeted attacks, the iOTN falls apart at a faster rate. Different targeted attacks have roughly the same influence on the network structure, but the destruction speed of the iOTN is different. In general, when the proportion of attacks on the economies reaches 40\%-60\%, the network will be close to collapse. The global oil trade network is a scale-free network, and a small number of economies have a large number of trade relationships \cite{Liu-Cao-Liu-Shi-Cheng-Liu-2020-Energy}. When the economies with more trade relations and trade influence are attacked, the structure of the iOTN will be more severely damaged.

It can be seen from the results of any simulated attacks that the iOTN has become more vulnerable to attacks over time. In other words, the iOTN becomes more fragile. Although the results in different years under the same attack strategy are not much different, the number of economies in the GCC decreases at a  faster rate. The structure and function of the iOTN are losing faster.

\begin{figure}[!t]
\centering
\hskip 0.6cm
\includegraphics[width=0.6\linewidth]{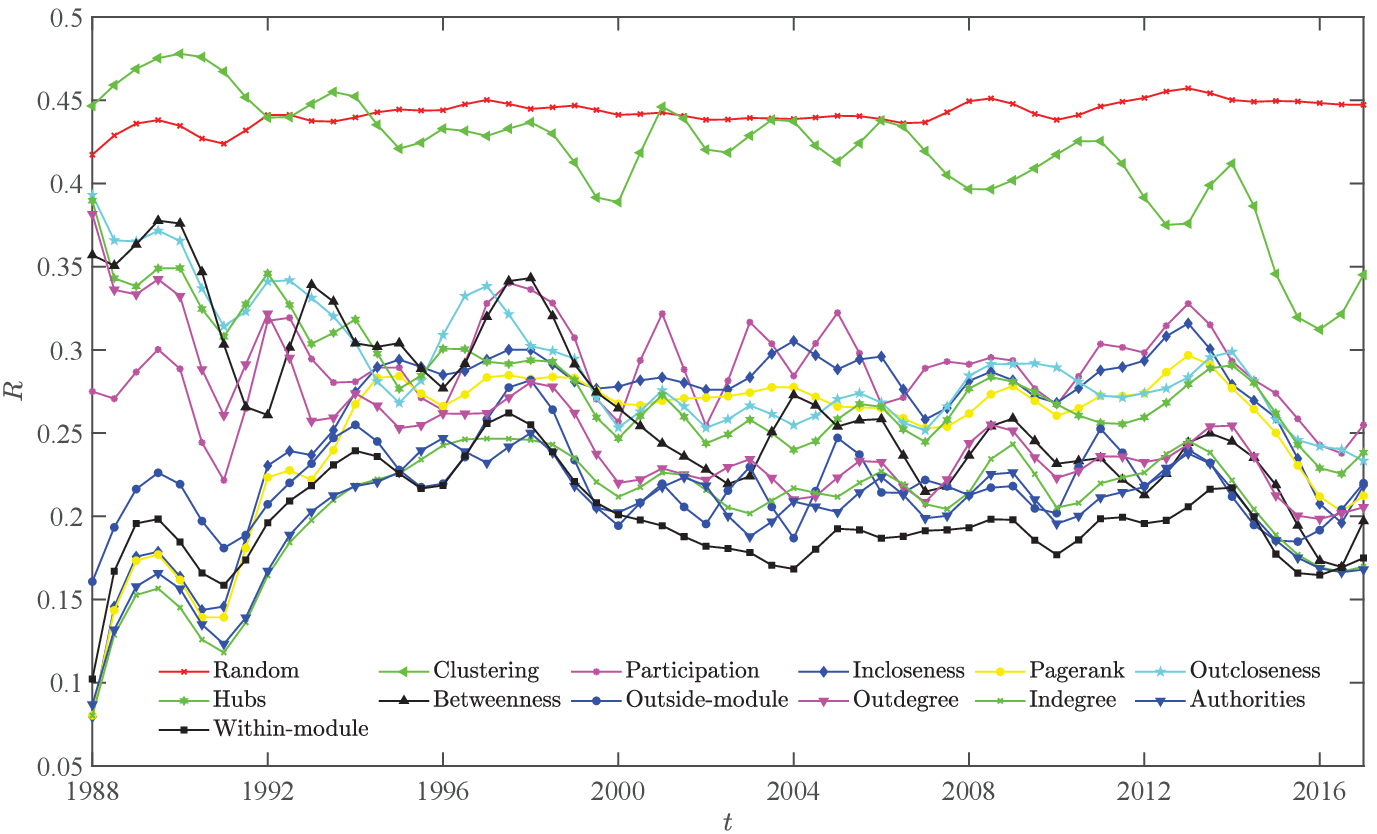}
\includegraphics[width=0.65\linewidth]{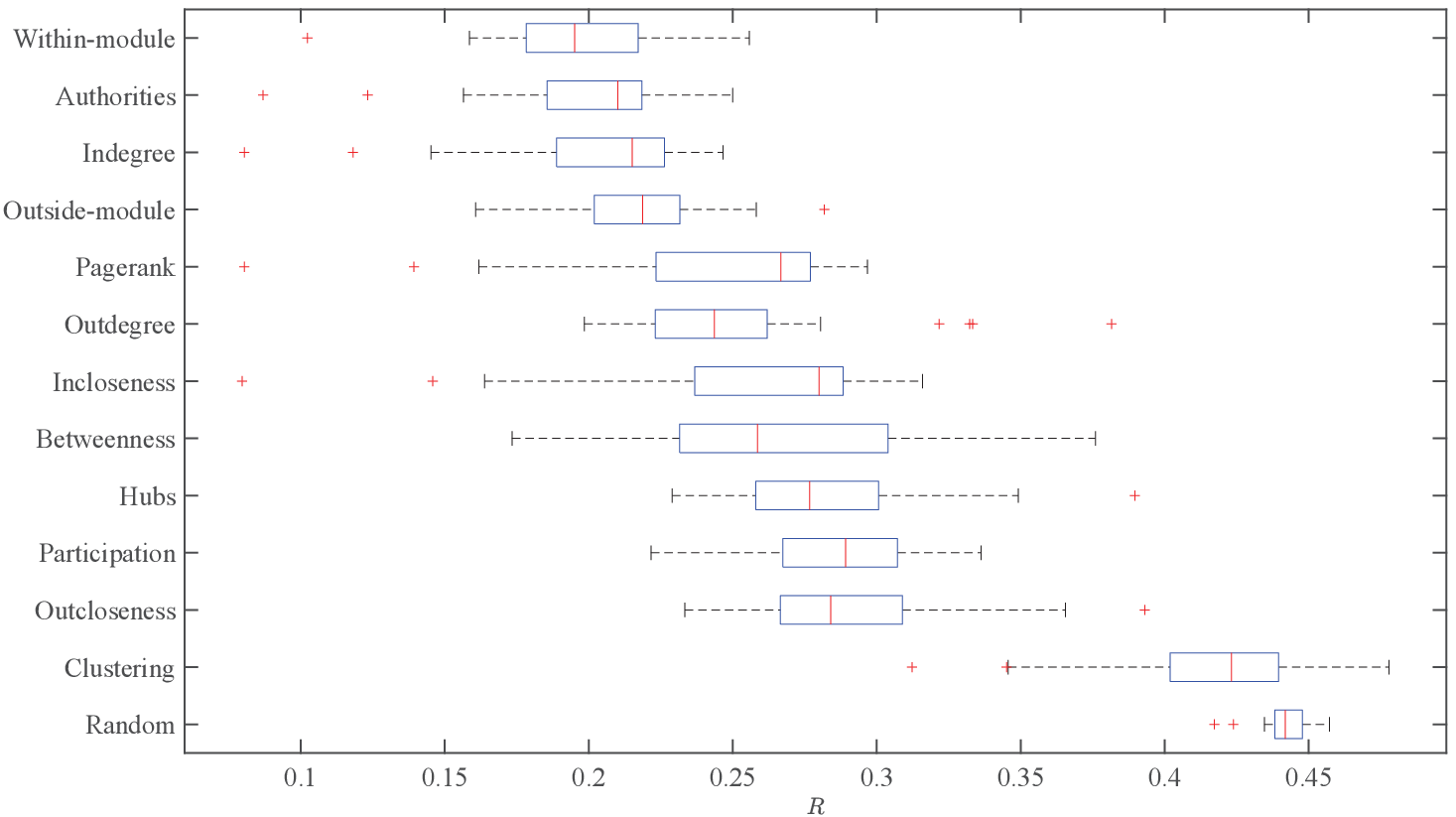}
\caption{The robustness evolution (Upper) and box plot (Lower) of the iOTN from 1988 to 2017.}
\label{Fig:iOTN:CentrAttackEvolution}
\end{figure}

Figure~\ref{Fig:iOTN:CentrRoleAttack} shows the vulnerability of the iOTN under specific attack strategies. It is necessary to calculate the entire network's robustness to compare different attack strategies and find the critical economies that significantly impact network stability. The robustness considers the destruction of the network under under all attack situations. According to Eq.~(\ref{Eq:Robustness:Measure}), we calculate the robustness of the iOTN in different years under various strategic attacks. To make the curve smoother, we use the cubic spline interpolation method to process
the data. Its evolution over time is shown in Fig.~\ref{Fig:iOTN:CentrAttackEvolution}.

It can be seen from Fig.~\ref{Fig:iOTN:CentrAttackEvolution} that under the random and local clustering coefficient-based attack strategies, the stability of the iOTN is the strongest, basically maintained at 0.4-0.5. In most targeted attacks, the anti-risk ability of the iOTN is low, and the network robustness is basically below 0.3. In most years, the attack strategy based on the within-module degree indicator results in the lowest robustness of the network, and it is the most effective attack strategy.

The within-module degree measures an economy's ability to establish connections with other economies within its module. The economy with high within-module degree plays a vital role in the aggregation of oil trade and is a critical connector. The failure of the connector will cause the collapse of regional trade cooperation, which will result in the loss of the structure and function of the entire trade network. These results also indicate that the concentration of regional oil trade may result in higher fragility of network structure. Strengthening the control of the connectors in each module is of great significance to maintaining the stability of the entire oil trade.

To observe the robustness rankings of the iOTNs under different strategic attacks, we arrange the robustness values over the past 30 years by year and draw the box plot, as shown in Fig.~\ref{Fig:iOTN:CentrAttackEvolution}, to observe its distribution.From the 30-year time scale, it can be clearly seen that when the oil trading system is attacked based on the within-module degree, the network robustness index is distributed evenly and basically stable at the level of 0.2. The second thing needs to observe is the authority strategy attack based on the global structure of iOTN.

The structure of the iOTN is complex, and its formation is affected by many factors and cannot be designed according to subjective will. Therefore, protecting economies that are critical to system stability is an important way to maintain oil trade stability. The numerical simulation experiments are of great significance for maintaining oil security and avoiding oil trade crises and global economic crises caused by systemic risks. For global policy researchers, finding the right indicators is a key factor, and the differences between indicators make it difficult to make optimal decisions in the decision-making process. Through horizontal comparison, we can find the indicator with the most significant impact on network connectivity. Of course, there are also many indicators of network stability, which can be measured not only by the proportion of the GCC, but also by other network functional indicators. We can also use the same method in this article to perform numerical simulations, determine the most influential attack indicators, and make corresponding decision support.

\section{Discussion and application}
\label{S1:Conclude}

We incorporated the topology indicators of the iOTN based on complex network theory into the traditional oil security and trade risk assessment framework, exploring the impact on the structural stability of the iOTN when economies with different important positions are attacked. Through the horizontal comparison of traditional centrality measures, the basis for maintaining network robustness and trade stability can be found.

By analyzing the different centrality positions of economies and international organizations in the iOTNs, we find that most influence measures have significant correlations, but the results of identifying the trade influence of economies by different node centrality measures are actually quite different. Optimal node impact measures also need to be selected based on the specific problem to be solved. At the individual level, the position of the economies in the global oil trade network is constantly changing with the network structure. In 2017, the Netherlands surpassed the USA to occupy a more important position and have higher structural importance. At the organizational perspective, international organizations such as the OPEC and the OECD have increased their trade relationships with other economies in the iOTN. However, their overall influence has shown a downward trend in recent years.

The simulated attacks results show that the international oil trade system has become more vulnerable in recent years. The system has lower network stability when it is attacked by an economy based on the within-module degree. The agglomeration of regional oil trade is a major reason for the vulnerability of the iOTN. The oil trading system is also fraught with uncertainty, with trade protection, national strategies, economic sanctions and even war all potentially putting the trading network at risk. Maintaining the trade system's stability requires a focus on the key connectors in the modular structure of the iOTN. The impact of the economies with the most significant influence within the module will cause the most incredible damage to the trade system's structure. Therefore, more consideration need be given to the trade gathering areas in the iOTN in the future, focusing on the economies with high within-module degree such as the Netherlands, the USA, and China.

The stability and systematic risk in the iOTN can be explored and identified based on the change of network structure \cite{Zhong-An-Gao-Sun-2014-PA,Du-Dong-Wang-Zhao-Zhang-Vilela-Stanley-2019-Energy}. However, to reduce the possibility of the outbreak of systemic risks in the global oil trade network, it is necessary to prevent the spread of trade shocks and improve the stability of the iOTN. In recent years, the international oil trade center has moved to the Asia-Pacific region, and European economies seek to reduce their dependence on Russian oil imports. Brexit (Britain exiting from the EU) and the American shale oil revolution have significantly impacted the oil trade. These facts fully reflect that oil trade status and policy changes in critical economies can cause instability and even endanger energy security. Therefore, identifying and protecting the economy with a critical position is of great significance for stabilizing the oil trade and prevent trade risks.

We provide a framework for selecting critical trade economies and maintaining the stability of the global oil trade system. However, the paper still has certain limitations. Firstly, the risks faced by the oil trade system come from many aspects, but our research only focuses on the situations in which the network structure can resist unexpected events such as trade interruptions. Secondly, there are many indicators for measuring network stability and robustness, but the indicators we choose only consider the integrity of the network structure. There has not been a more comprehensive measure of the impact on the actual function of the network. In the future, more realistic factors can be used to expand the assessment framework for the stability of the oil trading system and introduce more mechanisms for economic trade interactions so as to better inform global economic development and oil security in conjunction with network topology.

\section*{Acknowledgment}

This work was supported by the Shanghai Outstanding Academic Leaders Plan and the Fundamental Research Funds for the Central Universities.

\section*{Data Availability}

Oil data sets related to this article can be found at https://comtrade.un.org/, an open-source online data repository.


\end{document}